\documentclass[aps,prd,twocolumn,superscriptaddress,floatfix,longbibliography,10pt]{revtex4-2}

\usepackage{graphicx}
\usepackage{dcolumn}
\usepackage{bm}
\usepackage{hyperref}

\usepackage{graphicx} 
\usepackage{wrapfig} 
\usepackage{subcaption} 
\usepackage{amsmath}
\usepackage{amssymb}

\usepackage{orcidlink} 

\usepackage{siunitx}
\DeclareSIUnit{\electronvolt}{eV}
\DeclareSIUnit{\electron}{$e^-$}
\DeclareSIUnit{\positron}{$e^+$}
\DeclareSIUnit{\photon}{$\gamma$}
\DeclareSIUnit{\events}{events}
\DeclareSIUnit{\bunch}{bunch}

\usepackage{enumitem}
\setlist[itemize]{itemsep=-0.4em}

\begin{document}

\title{Feasibility of Low-Energy True Muonium Photoproduction}

\author{Ivo Schulthess\orcidlink{0000-0002-5621-2462}}
\email[Corresponding author: ]{ivo.schulthess@ethz.ch}
\affiliation{Institute for Particle Physics and Astrophysics, ETH Zurich, 8093 Zurich, Switzerland}

\author{Benjamin Banto Oberhauser\orcidlink{0009-0006-4795-1008}}
\affiliation{Institute for Particle Physics and Astrophysics, ETH Zurich, 8093 Zurich, Switzerland}

\author{Paolo Crivelli\orcidlink{0000-0001-5430-9394}}
\email[Corresponding author: ]{crivelli@phys.ethz.ch}
\affiliation{Institute for Particle Physics and Astrophysics, ETH Zurich, 8093 Zurich, Switzerland}

\date{\today}

\begin{abstract}
    True muonium, the bound state of a muon and an antimuon, is a theoretically well-understood but experimentally unobserved exotic atom. Its purely leptonic nature makes it a sensitive probe for bound-state quantum electrodynamics and possible physics beyond the Standard Model. We present a feasibility study of low-energy true muonium production via near-threshold photoproduction on a fixed target. The study includes simulations of the signal and dominant background processes, estimates of the required photon fluxes, and an overview of possible gamma sources. We show that the proposed Gamma Factory at CERN could provide the necessary photon energies and rates for the production of true muonium. Furthermore, cut-based selections are shown to suppress the background below the expected signal level. Finally, we discuss possible physics opportunities beyond the first observation, including measurements of its lifetime, hyperfine splitting, and Lamb shift.
\end{abstract}


\maketitle


\section{Introduction}\label{sec:intro}

True muonium (TM), the bound state of a muon and an antimuon ($\mu^+\mu^-$), is a uniquely clean system for precision studies of bound-state quantum electrodynamics (QED)~\cite{Lamm:2015fia, Lamm:2017ioi}. As an ``all-leptonic'' exotic atom, it is free of nuclear finite-size effects and is compact (Bohr radius $\sim 512~\mathrm{fm}$), which makes it simultaneously sensitive to higher-order QED corrections and to potential contributions from physics beyond the Standard Model. In particular, the muon anomalous magnetic moment provides one of the most stringent tests of the Standard Model in the lepton sector. Recent measurements by the Muon $g$--2 collaboration at Fermilab~\cite{Muong-2:2021ojo}, together with ongoing theoretical efforts to refine the Standard Model prediction~\cite{Keshavarzi:2022kpc}, have renewed interest in complementary probes of muon-sector interactions. A precise experimental handle on TM spectroscopy and decay observables would therefore be a valuable addition to the landscape of muon precision physics~\cite{Lamm:2015fia, Lamm:2017ioi, Karshenboim:2005iy}.

Despite considerable theoretical interest, TM has not yet been observed. Several experimental production concepts have been proposed, including threshold production in $e^+e^-$ collisions~\cite{Brodsky:2009gx, Bogomyagkov:2017uul, Fox:2021mdn}, fixed-target production with electrons or positrons on nuclei~\cite{Banburski:2012tk, Gargiulo:2024zyc, Agapov:2026wpq}, and searches at high-energy experiments such as LHCb~\cite{CidVidal:2019qub}. These approaches can offer discovery potential, but they typically yield TM at high or ill-defined kinetic energy, which complicates direct measurements of intrinsic properties such as the decay rate, hyperfine splitting, or Lamb shift.

\begin{figure}
    \centering
    \includegraphics[width=1.0\linewidth]{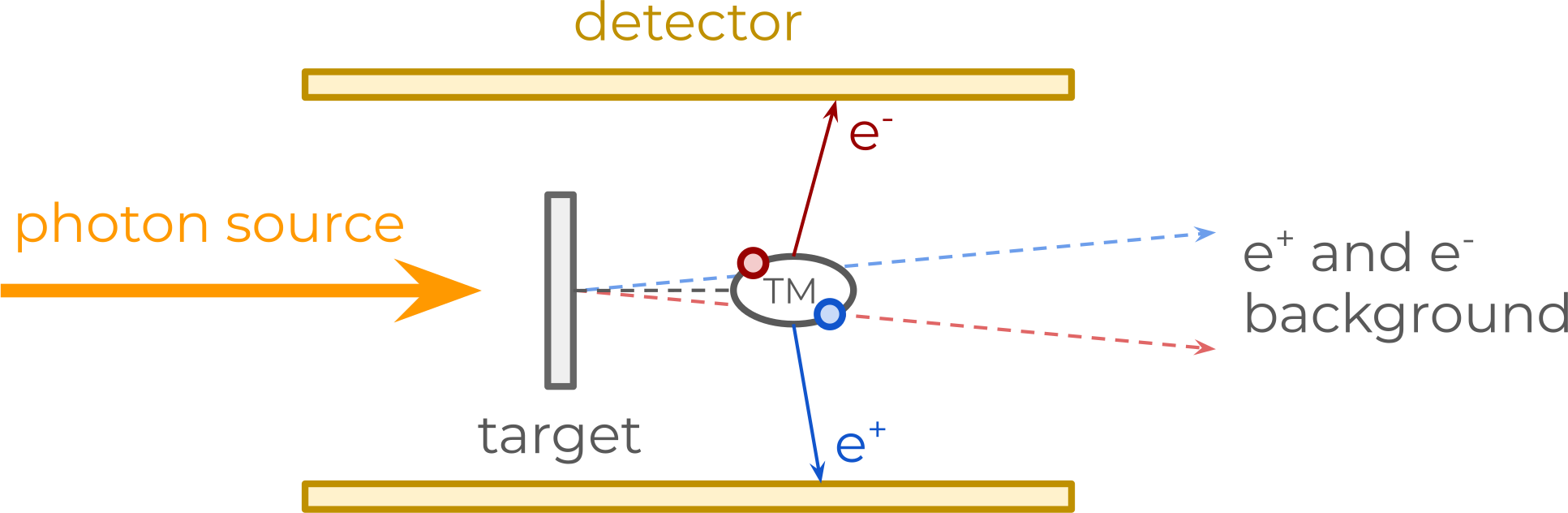}
    \caption{Schematic of the proposed method. Photons impinge on a lead target where lepton pairs can be produced. The muon pairs can form the bound state of interest, TM in the ortho state, which decays into an electron--positron pair. The background process in this setup is electron--positron pairs produced from other mechanisms, which are mostly boosted in the forward direction. }
    \label{fig:schematic}
\end{figure}

In this work, we explore an alternative route based on near-threshold photoproduction, $\gamma Z \to (\mu^+\mu^-)Z$, using photons with energies just above the production threshold $\gtrsim \SI{211}{\mega\electronvolt}$. In this regime, TM can be produced with low kinetic energy and can emerge from a thin target, opening a path towards controlled measurements of its properties. A further advantage is the distinct event topology: TM decays to an $e^+e^-$ pair on a timescale of order \SI{1.8}{\pico\second}, with the decay products nearly back-to-back in the low-energy limit, while the dominant QED backgrounds from direct pair production are strongly forward-peaked and can be reduced by geometric acceptance and kinematic selections. A schematic of the proposed method is presented in Fig.~\ref{fig:schematic}. 

The central challenge is the photon source. The TM photoproduction cross section scales steeply with nuclear charge and is small even for high-$Z$ targets. For lead, one expects $\sigma_{\gamma Z} = \SI{1.8E-35}{\centi\meter\squared}$~\cite{Ginzburg:1998df, Arteaga-Romero:2000mwd}, implying that extremely large photon fluxes are required to reach observable yields. Moreover, deviations from a quasi-monoenergetic spectrum are critical. Photons below the threshold contribute only to backgrounds, whereas photons far above the threshold increase the TM kinetic energy and reduce the accessibility of low-energy observables. This makes joint optimization of photon spectrum, target material, and detector acceptance essential, and it directly links the feasibility of TM to the capabilities of emerging high-intensity photon sources~\cite{Howell:2020nob, korol_novel_2022}.

We therefore performed a feasibility study that investigates near-threshold TM production and dominant background mechanisms in simplified target and detector configurations, and we discussed promising photon-source concepts that could approach the required energy and flux regime. Candidate source classes include crystal radiators~\cite{Bandiera:2015giw, korol_novel_2022, Sytov:2023epo}, plasma- and laser-driven concepts~\cite{Sampath:2020dec, Zhu:2020tlq, korol_novel_2022}, and future high-rate concepts such as the Gamma Factory~\cite{PBC:2025sny, Budker:2020zer, Bieron:2021ojp}. On the simulation side, we build on established particle-transport and electromagnetic frameworks~\cite{GEANT4:2002zbu, Ivanchenko:2020wvu} and incorporate a dedicated treatment of TM production and dissociation. The results provide an initial assessment of whether low-energy TM production via photoproduction may become experimentally accessible in the near future.

\section{Simulation Setup}\label{sec:setup}

To assess the efficiency of the proposed detection scheme and the level of background, we performed simulations using Geant4 v11.3.0~\cite{GEANT4:2002zbu}, which includes muon-pair production up to the threshold in the upgraded electromagnetic physics subpackage~\cite{Ivanchenko:2020wvu}. The target was placed at the origin $z=0$. We use the true particle properties without applying detector smearing or reconstruction effects. To obtain these, a cylindrical virtual detector with a length of \SI{1500}{\milli\meter}, from $z=\SI{-500}{\milli\meter}$ to $z=\SI{1000}{\milli\meter}$, and an inner radius of \SI{150}{\milli\meter} was placed around the target. This longitudinal range is sufficient to capture the relevant events. If necessary, an end-cap detector can be placed in the forward direction to record all background events. A schematic of the setup is shown in Fig.~\ref{fig:schematic}. 

The choice of target geometry is driven by the production and dissociation probabilities of TM. The dissociation length of TM in a solid target is $l_\mathrm{diss} = \frac{1}{n \sigma}$ with $\sigma = Z^2 \times 1.3 \times 10^{-21}~\mathrm{mm^2}$, where $n$ is the target density (number of atoms per unit volume) and $Z$ is the nuclear charge number of the target material~\cite{Holvik:1986ty}. We chose a lead target thickness of \SI{6.95}{\micro\meter}, corresponding to $2l_\mathrm{diss}$. Thinner targets reduce TM production, whereas thicker targets lead to significantly more background events but almost no improvement in signal yield, because the number of additionally formed atoms cancels out with the number of atoms that dissociate. 

In the simulations, we consider photons whose energies follow a Gaussian distribution with ${E_\gamma = \SI{300(30)}{\mega\electronvolt}}$. For the signal simulation, the muon-pair production is simulated with Geant4 using cross-section biasing to increase the number of generated events. The produced muon pairs are then assumed to form the ortho-TM bound state. The subsequent decay into an electron--positron pair is Monte-Carlo sampled using the kinematic variables of the TM, and the hit positions in the detector are calculated analytically. For the background simulation, no biasing is applied and \SI{4e13} primary photons are simulated directly. 

\begin{figure}
    \centering
    \includegraphics[width=1\linewidth]{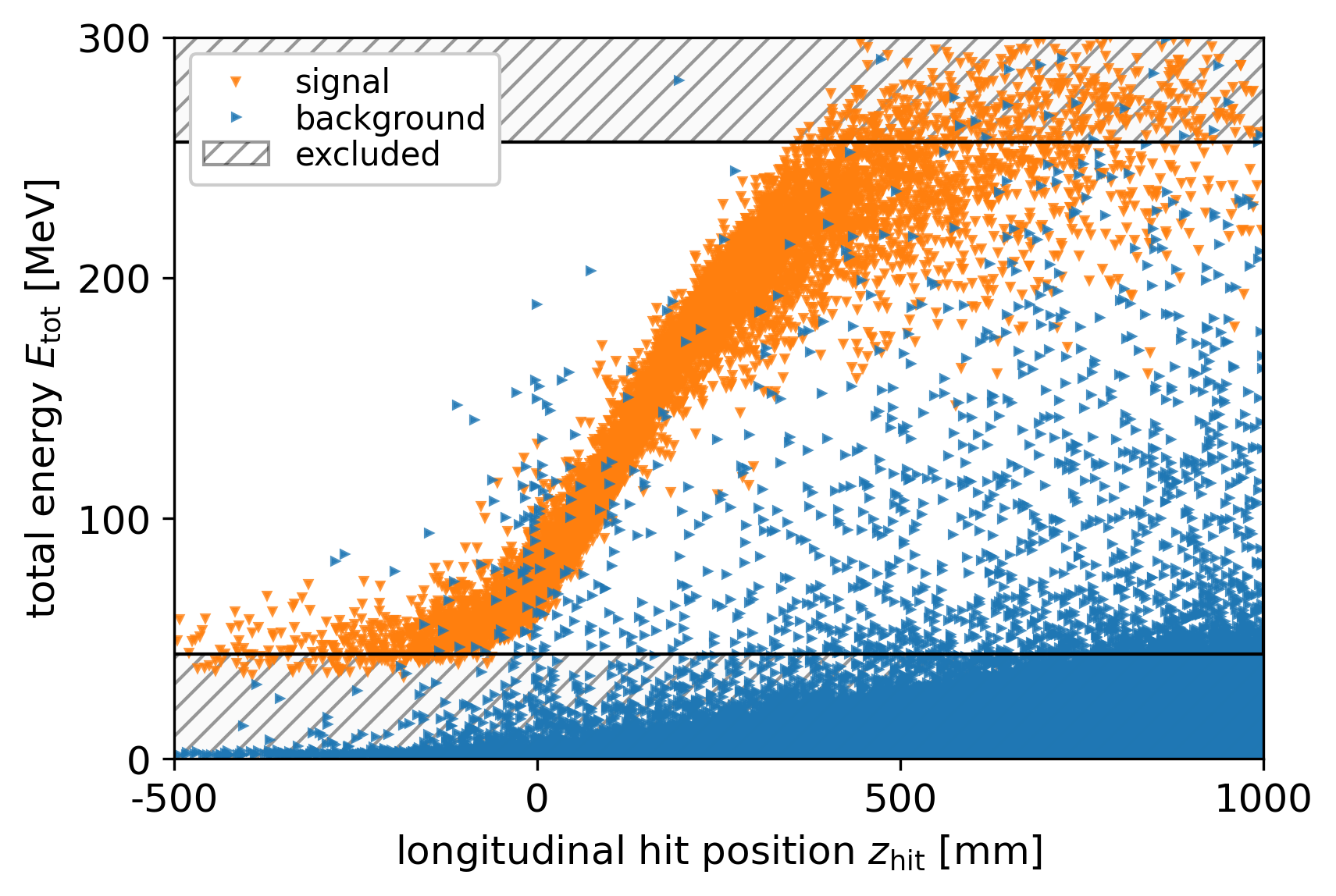}
    \caption{Event preselection for signal hits (orange) and background hits (blue) with cuts on the longitudinal hit position and the kinematically allowed energy range. }
    \label{fig:eventPreselection}
\end{figure}

We preselect only events within the kinematically allowed regime $E_\mathrm{th} = \gamma_\mathrm{TM} \, \left(m_\mu \pm \beta_\mathrm{TM} \sqrt{m_\mu^2 - m_e^2}\right)$, where $m_e$ is the electron mass, $\gamma_\mathrm{TM}$ and $\beta_\mathrm{TM}$ are the Lorentz gamma and beta factors of the TM, respectively. This leads to a range of $E_{\mathrm{tot}} \in [43.5,~256.5]$~MeV, which reduces the background events by a factor of \SI{5.9e4} and leads to $4.6 \times 10^{-8}~\mathrm{e^-/\gamma}$ and $4.1 \times 10^{-8}~\mathrm{e^+/\gamma}$. At the same time, 84.5\% of signal events are accepted by this preselection. The distribution of the longitudinal hit position $z_\mathrm{hit}$ and the total energy $E_\mathrm{tot}$ for signal and background hits is presented in Fig.~\ref{fig:eventPreselection}, highlighting the selected energy range. To further optimize the preselection, the acceptance region could be adjusted by introducing a position-dependent energy threshold to follow the signal event distribution.

\section{Photoproduction}\label{sec:photoproduction}

Various mechanisms exist for the photoproduction of muons. They include Bethe--Heitler ($\gamma+Z \to \mu^+ + \mu^- + Z$), triplet ($\gamma + e^- \to \mu^+ + \mu^- + e^-$), photon fusion ($\gamma + \gamma^* \to \mu^+ + \mu^-$), and meson production/decay. For near-threshold production, the first is expected to be dominant~\cite{Krass:1965hh}. The others are either forbidden or highly suppressed. To date, no cross-section calculation for near-threshold TM production has been carried out, and input from the theoretical community is required to obtain a more refined estimate of the production rate. 

In the high-energy limit $E_\gamma \gg 2 m_\mu$, the ortho-TM photoproduction cross section approaches~\cite{Ginzburg:1998df}
\begin{equation}\label{eq:oTMxSec}
    \sigma_{\gamma Z} = \SI{0.139}{\giga\electronvolt} \, \frac{\pi \alpha^8}{m_\mu^{4}} \, \frac{Z^4}{A^{2/3}} \, ,
\end{equation}
where $m_\mu$ is the muon mass, $\alpha$ the fine-structure constant, and $A$ the atomic weight. 

\begin{figure*}[!t]
    \centering
    \includegraphics[width=0.9\linewidth]{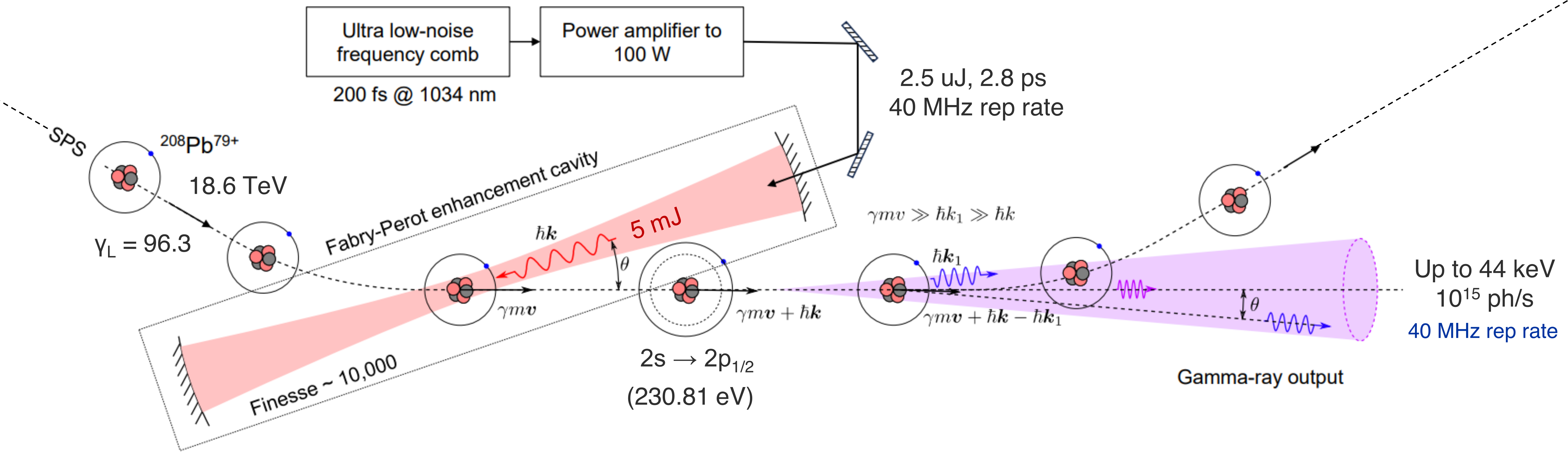}
    \caption{Schematic of the proof-of-principle setup of the Gamma Factory at CERN's SPS. Figure adapted from Ref.~\cite{Granados:2024sht}. Detailed description in the text. }
    \label{fig:gammaFactorySchematicSPS}
\end{figure*}

We computed the suppression factor $N_\mathrm{TM}/N_{\mu^+\mu^-}$ in the high-energy limit and obtained \SI{1.48E-8} at \SI{100}{\giga\electronvolt}. In the absence of dedicated near‑threshold calculations, we assume that this suppression factor is approximately energy independent and thus take the TM photoproduction rate to scale proportionally with the Bethe–Heitler muon-pair production rate down to threshold. A similar proportionality between bound‑state and continuum production is reported in $e^+e^-$-annihilation production process at high energy~\cite{Brodsky:2009gx}. This extrapolation should be viewed as a phenomenological estimate, since the suppression factor close to threshold has not yet been calculated and may differ substantially from the high-energy behavior. Using these assumptions for the TM photoproduction rate, we find that \SI{1.44E19} photons are required to produce one TM atom, which corresponds to a rate of \SI{6.97E-20} TM atoms per photon. The distributions of the electron--positron events from the TM decay are shown and compared to the background distributions in Sec.~\ref{sec:background}.

\section{Gamma Sources}\label{sec:sources}

To date, no gamma source exists that can provide photons at the required energy and flux. However, a few promising avenues exist, including CERN's Gamma Factory proposal, crystal radiators, and plasma sources. In what follows, we elaborate on these three options and estimate the TM production rate for the Gamma Factory. 

\subsection{Gamma Factory}

The Gamma Factory is a proposed facility at CERN that makes use of the Large Hadron Collider (LHC). Highly charged ions, such as hydrogen-like lead, Pb$^{81+}$, are accelerated to relativistic energy. The ions can be excited from the ground state with an optical laser, which is possible because in the ion's rest frame, the laser photon energy $E_l$ is increased by a factor $2\gamma$. The subsequent isotropic de-excitation of the ions leads to the emission of a photon that, in the laboratory frame, is boosted by another factor $2\gamma$, leading to a total enhancement of $E_\gamma = 4 \gamma^2 E_l$. In such a configuration, photon energies up to \SI{400}{\mega\electronvolt} are accessible with unprecedented rates of more than \SI{e16}{photons/s}~\cite{Krasny:2018xxv, Budker:2020zer}. A proof-of-principle experiment is planned at CERN's Super Proton Synchrotron (SPS) and aims to produce \SI{e15} photons per second with an energy of \SI{44}{\kilo\electronvolt}~\cite{Krasny:2019wch, Granados:2024sht}. A schematic of the concept is presented in Fig.~\ref{fig:gammaFactorySchematicSPS}. 

To produce gammas at an energy of $E_\gamma \approx \SI{300}{\mega\electronvolt}$, an atomic transition of about \SI{50}{\kilo\electronvolt} is required at the highest Lorentz boost of the ion of $\gamma \approx 3000$, limited by the magnetic rigidity of the LHC ring. As the 1s-2p transition scales as $\Delta E_\mathrm{1S-2p} \approx Z^2 \times \SI{10.2}{\electronvolt}$, the best suited ion is $^{174}_{70}\mathrm{Yb}^{69+}$. However, it requires a laser photon energy of $E_l = \SI{8.33}{\electronvolt}$ or a wavelength of \SI{149}{\nano\meter}. Although mirrors at this wavelength exist, they do not have the maturity yet to build high-power enhancement cavities~\cite{wilbrandt_protected_2014}.
\footnote{An alternative approach could be to use a laser with a wavelength of \SI{300}{\nano\meter}, excite the 1s-2s transition via two-photon excitation, and use a strong dipole to quench the 2s-2p transition.} 

\begin{figure*}[!tb]
    \centering
    \includegraphics[width=0.9\linewidth]{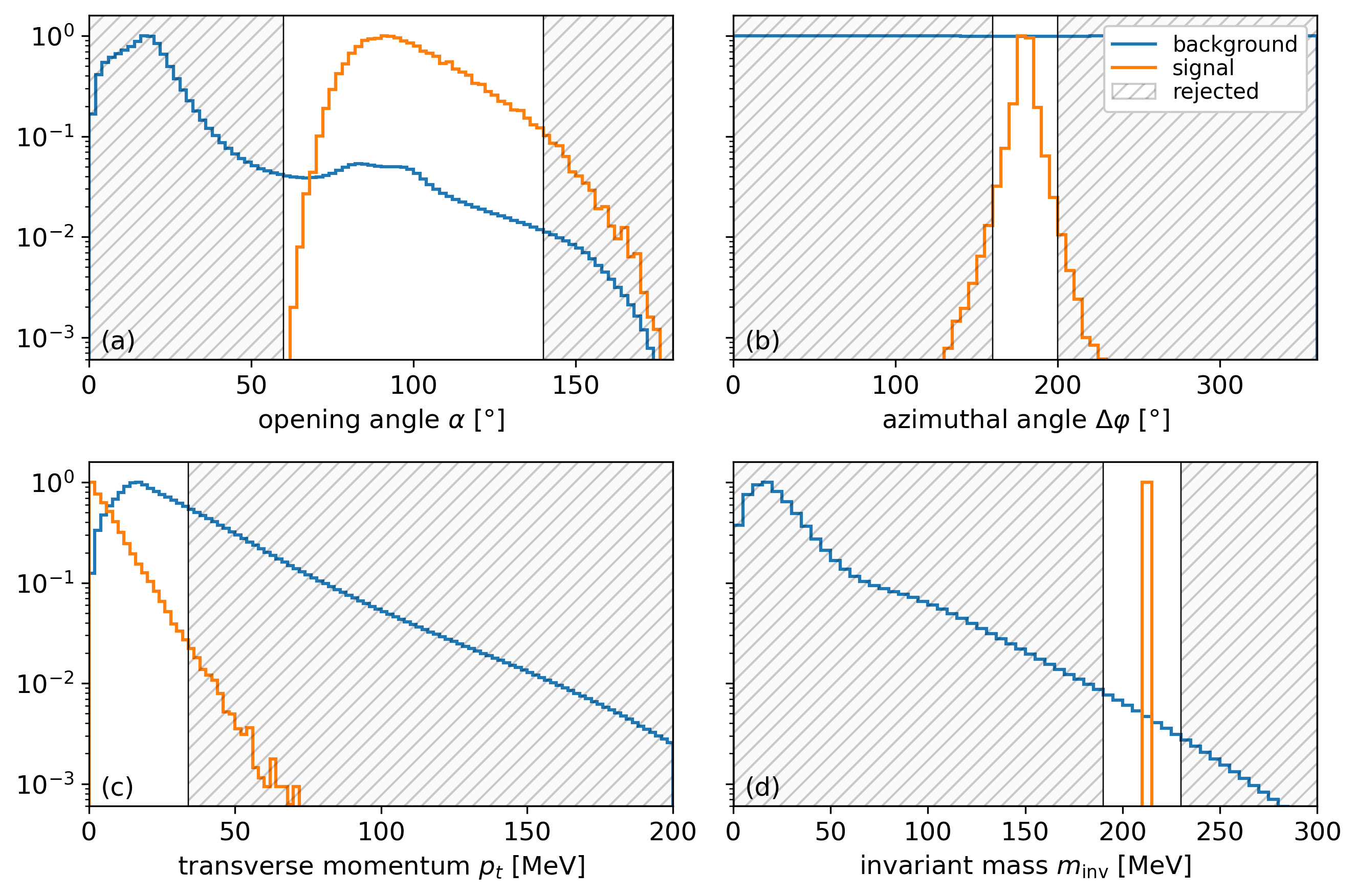}
    \caption{Distributions of signal (orange) and background events (blue) for (a) the opening angle and (b) the azimuthal angle difference between the electron and positron tracks, (c) the transverse momentum of the pair, and (d) the reconstructed invariant mass. The hatched areas indicate the rejected events. }
    \label{fig:eventDistributions}
\end{figure*}

Since the emission of the photons in the ion's rest frame is isotropic, the high-energy photons are in a forward cone with a small angle $\approx 1/\gamma$ and can be selected by collimation~\cite{Budker:2020zer}. Allowing a relative energy spread of 10\% will reduce the total flux by the same amount. 

For Pb$^{81+}$, a photon emission rate of \SI{3.8E8} photons per bunch was estimated, for a \SI{5}{\milli\joule} laser pulse of \SI{500}{\pico\second} duration~\cite{Bieron:2021ojp}. This leads to \SI{3.8E14} photons per second with 10\% energy spread at the planned bunch rate of \SI{10}{\mega\hertz}. The photon yield is proportional to the spontaneous decay rate, which is proportional to $Z^4$~\cite{Omidvar:1983rbq}. So, by changing the ion from Pb$^{81+}$ to Yb$^{69+}$, the decay rate is reduced by roughly a factor of two. Together with our cross-section estimate, this leads to a total period of \SI{1.4e11}{\bunch\per TM} or a rate of about \SI{1}{TM/d} at the proposed Gamma Factory. However, in this ``weak-saturation regime'', the photon yield scales linearly with the laser pulse length and energy, providing a direct path towards higher production rates with more powerful laser systems~\cite{Bieron:2021ojp}. 

\subsection{Crystal Radiators}
The strong crystalline fields of order \SI{e10}{\volt\per\centi\meter}, corresponding to \SI{3000}{\tesla}, enable the production of gamma radiation from charged particles via the coherent bremsstrahlung and channeling mechanism. Additionally, a (periodic) bending of the crystal can be introduced to alter the emission spectrum. At the desired photon energy of \SI{300}{\mega\electronvolt}, crystal radiators may even provide a flux exceeding that of the proposed Gamma Factory~\cite{korol_novel_2022}. Detailed configurations and simulations have to be performed using the channeling physics implemented in Geant4~\cite{Sytov:2023epo}. 

\subsection{Plasma Sources}
A third promising way to obtain a sufficiently high photon flux is through plasma sources. Although not yet mature enough, the field is rapidly developing, and various configurations have been suggested to provide high rates of photons in the regime of a few hundred MeV~\cite{Sampath:2020dec, Zhu:2020tlq, korol_novel_2022}.

\section{Background Rejection}\label{sec:background}

Although most of the electron--positron pairs from the Bethe--Heitler pair production in the thin Pb target are boosted in the forward direction, there are some hits around the location of the target with high energy, which potentially could mimic a signal event. We identified two main channels. Photons can cause a photonuclear reaction in the lead target, thus producing $\pi^0$
\begin{equation*}
    \gamma + \mathrm{Pb} \to \pi^0 + X \, ,
\end{equation*}
where $X$ denotes the hadronic nuclear remnant(s). The $\pi^0$ has two main decay channels that both contribute:
\begin{align*}
    \pi^0 &\to \gamma_1 + \gamma_2 \, , \quad \gamma_i + \mathrm{Pb} \to e^+e^- + \mathrm{Pb} \, \quad (i = 1~\mathrm{or}~2) \\
    \pi^0 &\to \gamma + e^+e^- \, .
\end{align*}
The first is dominant, but requires the subsequent conversion of one of the photons into an electron--positron pair. The second is the Dalitz decay, which produces an electron--positron pair directly at the decay vertex.

To discriminate signal events from background contributions, we apply a cut-based event selection strategy. Foremost is the timing acceptance, which is the time difference between electron and positron hits. As the Gamma Factory operates in bunches, all hits within the window of one bunch, corresponding to \SI{100}{\nano\second}, are accepted. In the case of a continuous source, such as a crystal radiator, time coincidence has to be applied. In what follows, we assume that electrons can be distinguished from positrons and that all other particles are either not detected or can be excluded. The Gamma Factory produces \SI{e8} photons per bunch, resulting in an average of \SI{4.6} and \SI{4.1} electrons and positrons per bunch after preselection, respectively. As they come in a single burst, this requires reconstructing all possible combinations, leading to about 19 pair candidates per bunch, or \SI{2.7e12} background events per signal event. 

Our event selection is based on four observables, which are the opening angle and azimuthal angle difference between the electron and positron tracks, the transverse momentum of the pair, and the reconstructed invariant mass, defined as
\begin{equation}\label{eq:alpha}
    \alpha = \arccos \left( \frac{\vec{p}_{e^+} \cdot \vec{p}_{e^-}}{|\vec{p}_{e^+}| \, |\vec{p}_{e^-}|} \right) \, ,
\end{equation}
\begin{equation}\label{eq:dphi}
    \Delta\varphi = \varphi_{e^+} - \varphi_{e^-} \, ,
\end{equation}
where $\phi_{e^\pm} = \arctan \left(p_{e^\pm}^y \, / \, p_{e^\pm}^x \right) $, 
\begin{equation}\label{eq:pt}
    p_t = \sqrt{\left( p_{e^+}^x +p_{e^-}^x \right)^2 + \left( p_{e^+}^y +p_{e^-}^y \right)^2} \, ,
\end{equation}
and
\begin{equation}\label{eq:mInv}
    m_\mathrm{inv} = \sqrt{\left( E_{e^+} + E_{e^-}\right)^2 - \left| \vec{p}_{e^+} + \vec{p}_{e^-} \right|^2} \, ,
\end{equation}
respectively. $E_{e^\pm}$ and $\vec{p}_{e^\pm}$ are the electron's and positron's energy and momentum vectors. The distributions of these observables for signal and background events are shown in Fig.~\ref{fig:eventDistributions}. 

Applying the following cuts allowed us to reject all background events in our simulated sample:
\begin{itemize}
    \item $\SI{60}{\degree} < \alpha < \SI{140}{\degree}$
    \item $\SI{160}{\degree} < \Delta\varphi < \SI{200}{\degree}$
    \item $p_t < \SI{34}{\mega\electronvolt}$
    \item $\SI{190}{\mega\electronvolt} < m_\mathrm{inv} < \SI{230}{\mega\electronvolt}$
\end{itemize}
This leads to a background efficiency of \textless \SI{9.8e-13} at 95\% C.L., while keeping the signal efficiency at 94.4\%. Since we estimated \SI{2.7e12} background events per signal event, this demonstrates that the background can be suppressed to a level significantly below the expected signal yield. 

We point out that the smearing of observables induced by a realistic detector response will substantially complicate background rejection. However, so far only four simple observables have been chosen, and very loose bounds have been applied. Other observables, such as the closest distance of approach, could also be considered. In combination with stricter cuts and more sophisticated selection methods, these approaches provide substantial potential for further improvement if required.


\section{Precision Measurements with True Muonium}

\begin{figure}
    \centering
    \includegraphics[width=1\linewidth]{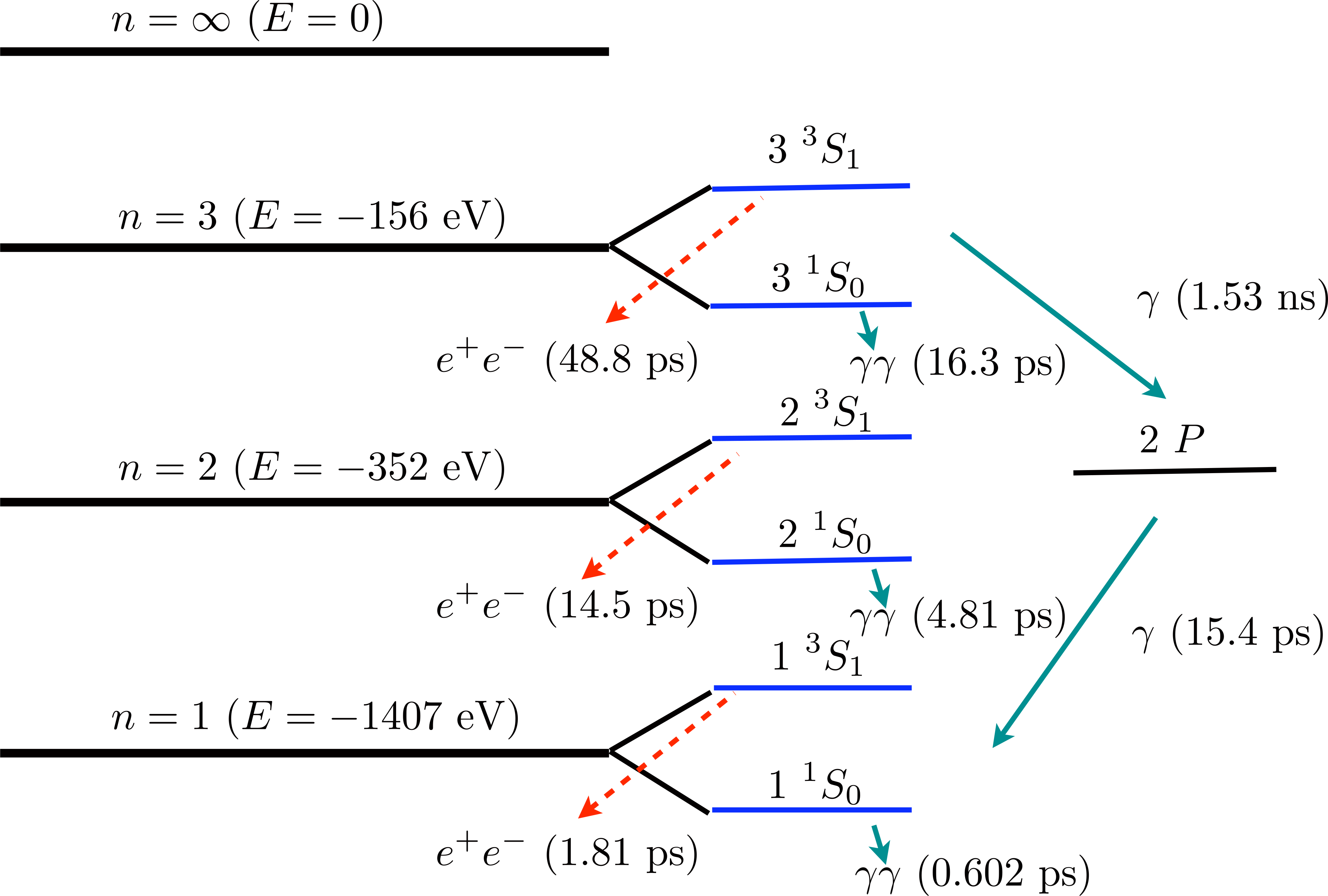}
    \caption{TM level diagram from Ref.~\cite{Brodsky:2009gx} (spacings not to scale). }
    \label{fig:Brodsky2009_fig1}
\end{figure}

Creating TM near threshold via photoproduction poses the advantage of having the atom almost at rest with a narrow energy distribution. This allows for not only its observations, but also the investigation of its properties, such as the lifetime, hyperfine splitting, or the Lamb shift. Figure~\ref{fig:Brodsky2009_fig1} shows the TM level diagram. The hyperfine splitting is the $1^1S_0 - 1^3S_1$ transition between the parallel and anti-parallel spin states, and the Lamb shift is the $2S_{1/2} - 2P_{1/2}$ transition. 

The observation of TM appears feasible with only a small number of events. The studies presented in Sec.~\ref{sec:background} show that the background can be efficiently suppressed below the expected signal level. A more precise estimate of the event yield required for a $5\sigma$ discovery will require a dedicated detector-level study, including residual backgrounds and reconstruction effects.

\begin{figure}
    \centering
    \includegraphics[width=1\linewidth]{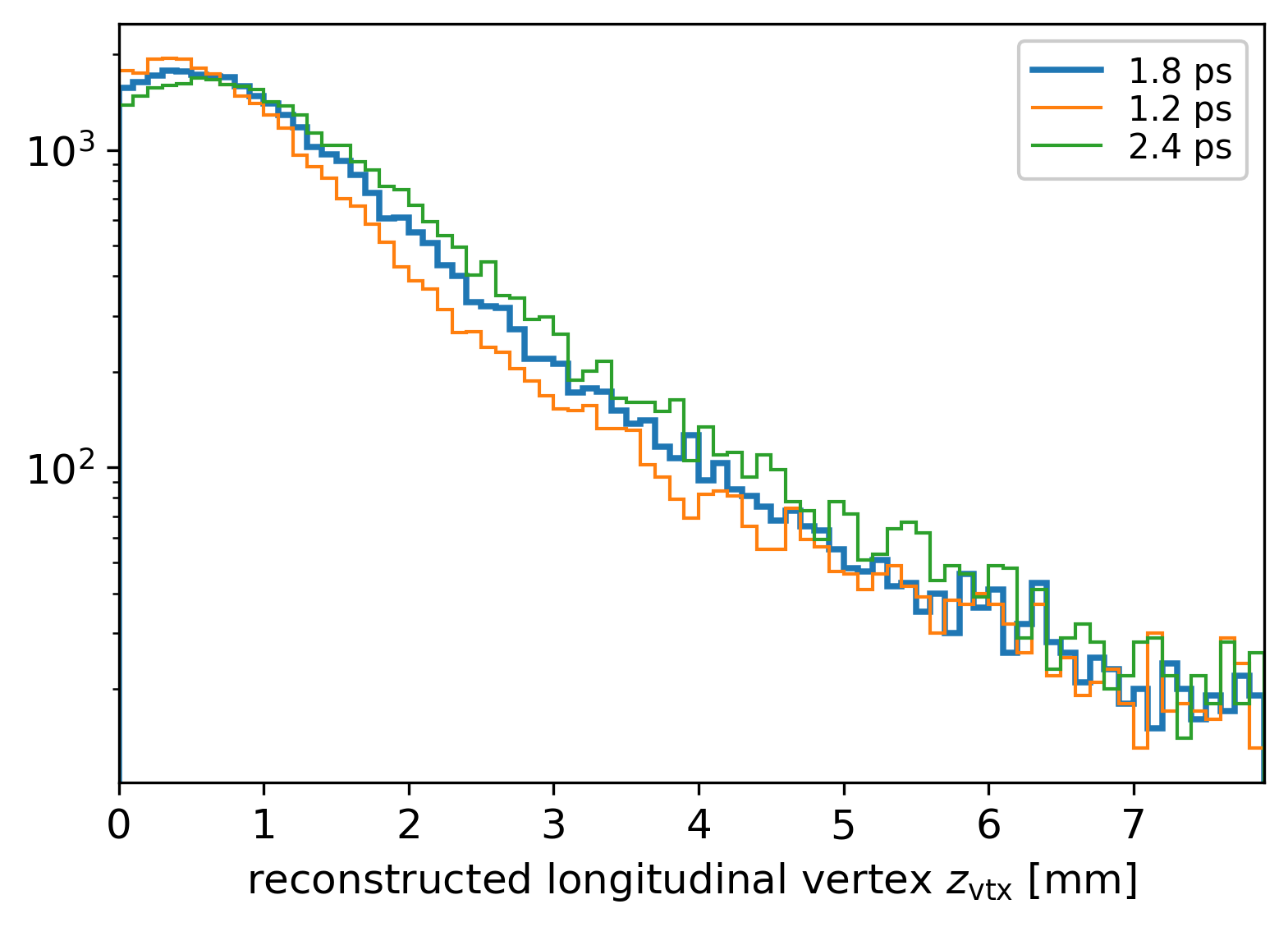}
    \caption{Distribution of \SI{5e4}{\events} of the reconstructed longitudinal vertex of the TM decay for the theoretical lifetime of \SI{1.8}{\pico\second} (blue) as well as a lower lifetime of \SI{1.2}{\pico\second} (orange) and a higher lifetime of \SI{2.4}{\pico\second} (green). }
    \label{fig:lifetime}
\end{figure}

After ortho-TM production is established, its lifetime of \SI{1.81}{\pico\second} can be measured by reconstructing the decay length or longitudinal decay vertex. For the energies considered here, the decay length is on the order of a few millimeters. Precise tracking of the electron--positron pairs is required. A detector concept similar to the Mu3e experiment, featuring an ultra-low material budget and excellent spatial and angular resolution for low-momentum electrons and positrons, is ideally suited for this purpose~\cite{Mu3e:2020gyw, Loreti:2024jkf}. Once realistic detector effects are included, the reconstructed vertex distributions deviate significantly from a pure exponential shape. The lifetime must therefore be extracted by comparing the reconstructed distributions to detector-level Monte Carlo predictions generated for different lifetime hypotheses. Three example distributions corresponding to different assumed TM lifetimes are shown in Fig.~\ref{fig:lifetime}. Detector effects were approximated through a simplified smearing model.

Beyond observing TM and determining its lifetime, its atomic characteristics, such as the Lamb shift and hyperfine splitting, can be explored through spectroscopic measurements. While theory predicts a transition frequency of about \SI{10}{\tera\hertz} for the Lamb shift, the hyperfine splitting transition of \SI{42}{\tera\hertz} is accessible today via \SI{7}{\micro\meter} quantum cascade lasers~\cite{Lamm:2017ioi}. The measurement principle is analogous to that applied in the muonium ($\mu^+e^-$) Lamb shift measurement~\cite{Mu-MASS:2021uou}. The laser frequency can be scanned over the $1^1S_0 - 1^3S_1$ transition. When on resonance, the electron--positron signal disappears since the excited state decays into two photons. Given that the decay length is only a few millimeters, as shown in Fig.~\ref{fig:lifetime}, the laser can be easily optimized to encompass the region in which TM is present.

\section{Conclusion}\label{sec:conclusion}

We presented a feasibility study of low-energy TM production via near-threshold photoproduction. Using Geant4 simulations, we studied the signal topology and dominant background processes. The results show that the background can be suppressed below the expected signal level using simple kinematic selections, suggesting that only a small number of events may already be sufficient for the observation of TM.

We discussed several possible gamma-source concepts and identified the proposed Gamma Factory at CERN as a promising option. With the currently proposed parameters, production rates on the order of one TM atom per day appear achievable, and future improvements to laser systems could increase the rate.

Near-threshold production would enable not only the first observation of TM but also studies of its lifetime, hyperfine splitting, and Lamb shift. Further detector-level simulations, improved theoretical calculations of the near-threshold production cross section, and dedicated source studies will be required to refine these estimates.

\begin{acknowledgments}
We gratefully acknowledge D.~Banerjee, E.~Depero, L.~Molina Bueno, and H.~Sieber for their earlier contributions to true muonium production studies and related Geant4 developments relevant to this work. PC would like to acknowledge the very useful discussion with Witek Kransy. This work was supported by the Swiss National Science Foundation under grant no. 230596. 
\end{acknowledgments}

\bibliography{ref}

\end{document}